\documentclass[epj,nopacs,final]{svjour}
\usepackage{graphics}
\usepackage{latexsym}
\usepackage{cite}
\usepackage{subfigure,wrapfig,multirow}
\usepackage{epsfig,color,rotating,amsmath,delarray,array}
\usepackage{makeidx,pifont,float,amssymb}
\definecolor{gray01}{gray}{0.9}
\definecolor{gray02}{gray}{0.8}
\definecolor{gray03}{gray}{0.7}
\definecolor{gray04}{gray}{0.6}
\definecolor{gray05}{gray}{0.5}
\definecolor{gray06}{gray}{0.4}
\definecolor{gray07}{gray}{0.3}
\definecolor{gray08}{gray}{0.2}
\definecolor{gray09}{gray}{0.1}

\newcommand{\er}{$\pm$}
\newcommand{\bc}{\begin{center}}
\newcommand{\ec}{\end{center}}
\newcommand{\be}{\begin{equation}}
\newcommand{\ee}{\end{equation}}

\begin{document}

\title{Heavy-flavor baryons}
\titlerunning{Heavy-flavor baryons}
\authorrunning{E. Klempt and S. Neubert}
\author{Eberhard Klempt and Sebastian Neubert}

\institute{Helmholtz-Institut f\"ur Strahlen- und Kernphysik,
Universit\"at Bonn, Germany}

\abstract{This is a contribution to the review ``50 Years of Quantum Chromdynamics"
edited by F. Gross and E. Klempt, to be published in EPJC. The contribution reviews the
properties of baryons with one heavy flavor: the lifetimes
of ground states and the spectrum of excited states. The importance of symmetries
to understand the excitation spectrum is underlined. An overview of searches for
pentaquarks is given.}
\date{Received: \today / Revised version:}

\mail{klempt@hiskp.uni-bonn.de}

\maketitle

\section{Introduction}
Baryons with one heavy quark $Q$ and a light diquark $qq$ provide an
ideal place to study diquark correlations and the dynamics of the
light quarks in the environment of a heavy quark. The heavy quark is
almost static and provides the color source to the light quarks.
Here, we attempt to understand the dynamics leading to the spectrum
of baryons with one heavy quark.

The Review of Particle Physics~\cite{ParticleDataGroup:2022pth} lists 28
charmed bary\-ons (16 with known spin-parity) and 22 bottom baryons
(15 with known spin-parity). One doubly charmed state has been
detected, the ground state $\Xi_{cc} ^{++}$. (Its isospin partner
$\Xi_{cc} ^{+}$ is known as well, with poor evidence and one star in
RPP, but we do not count isospin partners separately.) In the decays
of the lightest bottom baryon, exotic $J/\psi p$ states,
incompatible with a three-quark configuration, have been have been
reported in studies of the reaction $\Lambda_b\to J/\psi p
K^-$~\cite{LHCb:2015yax,LHCb:2019kea}. The search for further states
and attempts to understand the underlying dynamics of heavy baryons
are active fields in particle physics. New information can be
expected from the upgrades of LHC, BELLE and J-PARC, and from the
new FAIR facility at GSI.

\section{Ground states of heavy baryons}

\subsection{Masses and lifetimes}
Table~\ref{B:mass-tau} presents masses and life times of the ground
states of heavy baryons containing a $b$-quark. Naively, one could
expect all these life times to represent the life time of the $b$
quark, that they all agree with the life time of the $B^0$ meson.
This life time is $\tau_{B^0}$=(1519\er 4)\,fs. Indeed, all life
times agree within $\sim 10$\% percent.

\begin{table}[t!]
    \caption{Masses and lifetimes of baryon ground states with one $b$-quark.
    The second line gives the mass in MeV, the third line the life time in fs.}
    \label{B:mass-tau}
\renewcommand{\arraystretch}{1.4}
    \begin{tabular}{cccc}
    \hline\hline
  $\Lambda_b ^0$   & $\Xi_b ^-$ & $\Xi_b ^0$ & $\Omega_b ^-$\\\hline
5619.60\er 0.17& 5797.0\er0.6& 5791.9\er 0.5& 6045.2\er 1.2\\
 1464\er 11  &  1572\er 40 & 1480\er 0.030 &1640$^{+180}_{-170}$\\
 \hline\hline
    \end{tabular}
\end{table}

This is not at all the case when the $b$-quark is replaced by a
$c$-quark (see Table~\ref{C:mass-tau}). The $D^0$ has a life time
$\tau_{D^0}=(410.3\pm 1.0)$\,fs, the  $D^+$ has
 $\tau_{D^+}=(1033\pm 5)$\,fs. The life times of charmed baryons are spread
 over a wide range and do not agree with the life times of $D$ mesons.
 In addition to the decay of the $c$-quark, the $c\bar d$ pair in a $D^0$ meson can annihilate
 into a $W^+$, a process forbidden for the $D^+$. In $B$ decays, the corresponding
CKM matrix element is small, and this effect is suppressed. Further
significant corrections are required to arrive at a consistent
picture for the decays of charmed mesons and baryons. The authors of
Ref.~\cite{Gratrex:2022xpm} have performed an extensive study of the
lifetimes within the heavy quark expansion, and have included all
known corrections. The impact of the charmed-quark mass and of the
wavefunctions of charmed hadrons were carefully studied. Then,
qualitative agreement between their calculations and the
experimental data was achieved.
\begin{table}[pt]
    \caption{Masses and lifetimes of baryon ground states with one $c$-quark.
    The second line gives the mass in MeV, the third line the life time in fs.}
    \label{C:mass-tau}
\renewcommand{\arraystretch}{1.4}
    \centering
    \begin{tabular}{cccc}
    \hline\hline
  $\Lambda_c ^0$  & $\Xi_c ^+$     & $\Xi_c ^0$ & $\Omega_c ^-$\\\hline
2286.46\er 0.14    \hspace{-2mm}&\hspace{-2mm}2467.71\er0.23 \hspace{-2mm}&\hspace{-2mm} 2470.44\er 0.2\hspace{-2mm}&\hspace{-2mm} 2695.2\er 1.7\\
 201.5\er 2.7  &  453\er 5    & 151.9\er 2.4 &268\er26\\
 \hline\hline
    \end{tabular}
\end{table}

\begin{figure*}[tt]
\begin{tabular}{ccc}
\includegraphics[width=0.32\textwidth,height=0.28\textwidth]{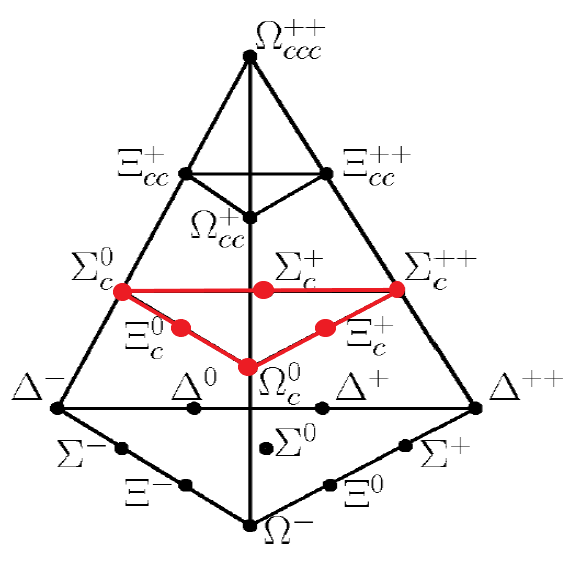}&
\includegraphics[width=0.32\textwidth]{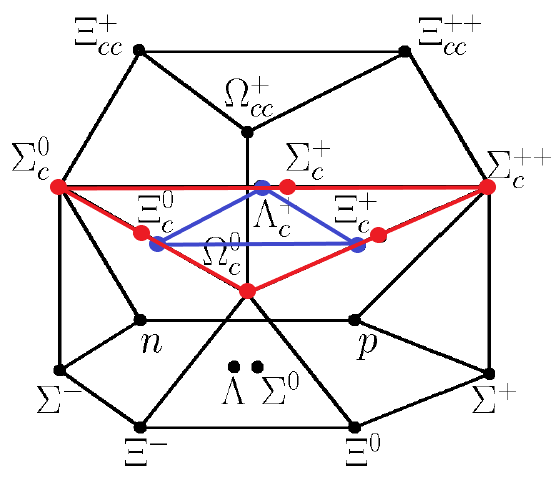}&
\raisebox{10mm}{\includegraphics[width=0.25\textwidth]{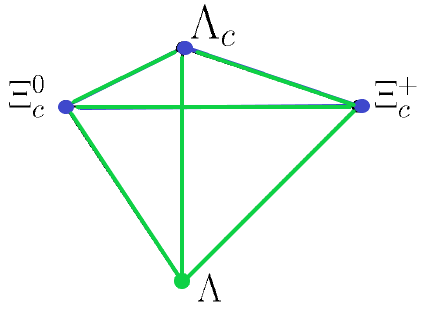}}\vspace{-1mm}
\end{tabular}
\caption{\label{fig:SU4a} Ground-state heavy baryons in SU(4).
Baryons with one charm quark are represented by colored dots. Left:
The symmetric 20-plet. Center: Baryons in the mixed-symmetry
20-plet. The mixed-symmetry 20-plet contains a sextet with a
symmetric light-quark pair (SU$_{\rm F}$(3) multiplicity 6) and a
triplet with an anti-symmetric light-quark pair (SU$_{\rm F}$(3)
multiplicity $\bar 3$). Right: The fully antisymmetric 4-plet. \vspace{-1mm}}
\end{figure*}
The first state with two charmed quarks, the $\Xi_{cc} ^{+}$ was
reported by the SELEX collaboration in two decay modes at a  mass of
(3518.9\er 0.9)\,MeV and with 5-6$\sigma$
\cite{SELEX:2002wqn,SELEX:2004lln}. In later searches, this state
was never confirmed. The LHCb collaboration found its doubly charged
partner $\Xi_{cc} ^{++}$ \cite{LHCb:2019epo}. Its mass is (3621.6\er
0.4)\,MeV, its life time (25.6\er 2.7)\,fs. Later, the LHCb
collaboration reported evidence for a $\Xi_{cc} ^{+}$ baryon at
(3623.0\er 1.4)\,MeV\,\cite{LHCb:2021eaf}. It is seen with
3-4$\sigma$ only but its mass is better compatible with an
interpretation of $\Xi_{cc} ^{+}$ and $\Xi_{cc} ^{++}$  as isospin
partners. A search for the $\Xi_{bc} ^{+}$ remained
unsuccessful~\cite{LHCb:2022fbu}.
\subsection{The flavor wave function: SU(4)}
In this contribution we discuss baryons with one heavy-quark flavor,
with either a charm or a bottom quark. Overall, we consider five
quarks, $u,d,s,c,b$, but we will not discuss bary\-ons with one
light ($q=u,d,s$) and two different heavy quarks like
$\Xi^+_{cb}=(ucb)$. Thus we can restrict ourselves to SU(4). The
four quarks have very different masses, and the SU(4) symmetry is
heavily broken, nevertheless it provides a guide to classify
heavy-quark baryons. Three-quark baryons can classified according to
\begin{equation}
4\otimes4\otimes4=20_{\rm s}\oplus20_{\rm m}\oplus20_{\rm
m}\oplus4_{\rm a}
\end{equation}
into a fully symmetric 20-plet, two 20-plets of mixed symmetry and a
fully antisymmetric 4-plet. In states with one heavy quark only,
there is one light quark pair. The light diquark can be decomposed
\begin{equation}
3\otimes3=\bar 3_{\rm a}\oplus6_{\rm s}
\end{equation}
The light diquark in the 6-plet is symmetric, in the $\bar 3$-plet
antisymmetric.

Figure~\ref{fig:SU4a}a shows the symmetric 20-plet, which contains
the well-known baryon decuplet and a sextet of charmed baryons. In
addition to  $\Xi_{cc} ^{+}$ and $\Xi_{cc} ^{++}$, a  $\Omega_{cc}
^{+}$ (with two charmed and one strange quarks) and a $\Omega_{ccc}
^{++}$ are expected but not yet observed. All baryons in the
symmetric 20-plet in the ground state have a total spin $J=3/2$. The
three quark pairs are symmetric with respect to (w.r.t.) their
exchange, in particular the pair of light quarks is symmetric w.r.t.
their exchange, they have SU$_{\rm F}$(3) multiplicity 6. Baryons
with three charmed quarks have not yet been discovered.

Figure~\ref{fig:SU4a}b shows the mixed symmetry 20-plet of heavy
bary\-ons. In the ground state they have $J=1/2$. Baryons with one
heavy quark occupy the second layer. The 6-plet and the $\bar
3$-plet are indicated. The sextet in the first floor has a a
symmetric light-quark pair, the two light-heavy quark pairs are then
antisymmetric in flavor. The 3-plet in the first floor has an
antisymmetric light-quark pair, the light-heavy quark pairs are then
symmetric in flavor.

Finally, there is a fully anti-symmetric 4-plet. It is shown in
Fig.~\ref{fig:SU4a}c. Ground-state baryons have a symmetric spatial
wave function. A spin wave function of three fermions has mixed
symmetry. A fully symmetry, a fully antisymmetric and a
mixed-symmetry wave function cannot be coupled to a fully symmetric
wave function. Hence ground-state baryons cannot be in the 4-plet.
Only excited baryons can have a fully antisymmetric flavor wave
function. Below, in Section~\ref{HBaTQS}, the wave functions and
their symmetries are discussed in more detail.

\section{Excited baryons: Selected experimental results}

\subsection{BaBar, BELLE and LHCb:}
Most information on heavy baryons stems from three experiments,
BaBar, BELLE and LHCb even though many discoveries had already been
made before with the Split-Field-Magnet, by the SELEX, UA and LEP
experiments at CERN, and by the CDF experiment at FERMILAB. BaBar at
SLAC (US) and BELLE at KEK (Japan) study the decays of $B$ mesons
produced in asymetric $e^+e^-$ storage rings with beam energies of 9
(KEK: 7)\,GeV for electrons and 3.1 (KEK: 4)\,GeV for positrons
resulting in a center-of-mass energy equal to the $\Upsilon(4S)$
mass of 10.58\,GeV. The LHCb experiment is placed at the Large
Hadron Collider at CERN operating at $\sqrt s=13.6$\,GeV. The
experiment is a single-arm forward spectrometer covering the
pseudorapidity range $2\leq\eta\leq5$. It is designed for the study
of particles containing $b$ or $c$ quarks. All three detectors have
vertex reconstruction capabilities; BaBar and BELLE track charged
particles in tracking chambers placed in the 1.5\,T magnetic field
of a superconducting solenoid. Particle identification is provided
by a measurement of the specific ionization and by detection of the
Cherenkov radiation in reflecting ring imaging Cherenkov detectors.
CsI(Tl)-crystal electromagnetic calorimeters allow for energy
measurements of electrons and photons. LHCb is equipped with
silicon-strip detector located upstream and downstream of a dipole
magnet with a bending power of about 4\,Tm. Photons, electrons and
hadrons are identified by a calori\-meter system consisting of
scintillating counters and pre-shower detectors, and an
electromagnetic and a ha\-dronic calorimeter. Muons are identified
by a system composed of alternating layers of iron and multiwire
proportional chambers.

In the following we discuss three important results from these
experiments that demonstrate the capabilities of the detectors.

\subsection{Observation of $\Omega_c ^{*0}(2770)$
decaying to $\Omega_c ^0 \gamma$ by BaBar:} 
\begin{figure}[pb]
\includegraphics[width=1\linewidth]{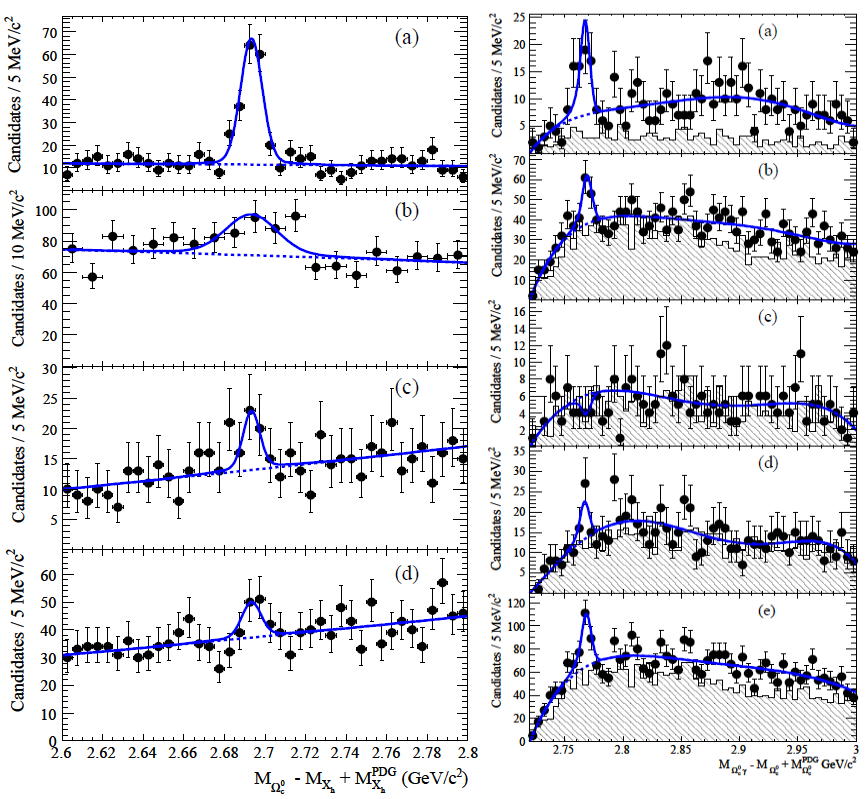}
\caption{\label{Babar}Left: The invariant mass distributions of
$\Omega_c ^0$ candidates in their decay to $\Omega^-\pi^+$ (a),
$\Omega^-\pi^+\pi^0$ (b), $\Omega^-\pi^+\pi^-\pi^+$ (c),
$\Xi^-K^-\pi^+\pi^+$ (d). $M_{\Omega_c ^0}$ is the reconstructed
mass of $\Omega_c ^0$ candidates, $X_h$ denotes the daughter
hyperon. Right: Invariant mass distribution of $\Omega_c
^*\to\Omega_c\gamma$ for the indivual $\Omega_c ^0$ decay modes
(a-d) and for the sum (e). (Adapted from
~\cite{BaBar:2006pve}.)\vspace{-3mm} }
\end{figure}
The Babar experiment
studied the inclusive reaction $e^+e^-\to \Omega_c ^{*0}\,X$ where
$X$ denote the recoiling particles~\cite{BaBar:2006pve}. $\Omega_c
^0$ baryons are identified via different decay modes and
reconstructed with a mass resolution $\sigma_{\rm RMS}=13$\,MeV. The
$\gamma$ is reconstructed in the $\Omega_c ^0$ CsI(Tl)
calori\-meter. Figure~\ref{Babar} shows the reconstructed $\Omega_c
^0$ and the $\Omega_c ^{*0}$ in its $\Omega_c ^{*0}\to\Omega_c
^0\gamma$ decay. Obviously, the $\Omega_c ^{*0}(2770)$ is equivalent
to $\Delta^0(1232)$ with the $u,d,d$ quarks exchanged by $c,s,s$,
and the transition corresponds to the $\Delta(1232)\to N\gamma$
decay.\vspace{-1mm}
\subsection{First determination of the spin and parity
of the charmed-strange baryon $\Xi_c ^+(2970)$ by BELLE.} 
\begin{figure}[ph]
\includegraphics[width=1\linewidth]{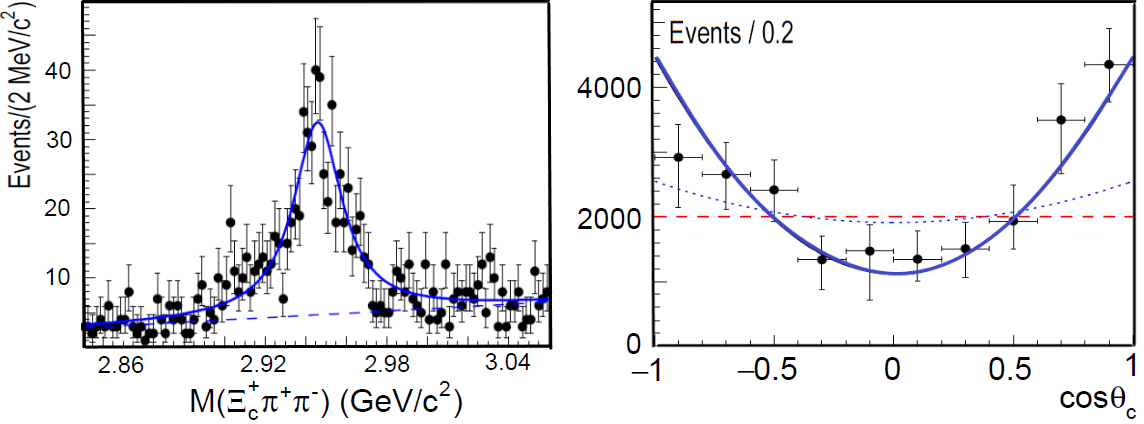}
\caption{\label{Belle}Left: The $\Xi_c ^+\pi^-\pi^+$ invariant mass
distribution for events in which the $\Xi_c ^+\pi^-$ invariant mass
is compatible with the $\Xi_c ^0(2645)$ mass. Right: The helicity
angle $\theta_c$ between the direction of the $\pi^-$ relative to
the opposite direction of the $\Xi_c ^+(2970)$ in the rest frame of
the $\Xi_c ^0(2645)$. (Adapted from Ref.~\cite{Belle:2020tom}.) \vspace{-9mm}}
\end{figure}

The BELLE
collaboration identified $\Xi_c ^+(2970)$ in the decay chain $\Xi_c
^+(2970)\to\Xi_c ^0(2645)\pi^+\to\Xi_c ^+\pi^-\pi^+$; $\Xi_c ^+$ is
reconstructed from its decay into
$\Xi^-\pi^+\pi^+$~\cite{Belle:2020tom}. Due to its mass, $\Xi_c
^0(2645)$ is likely the spin excitation with $J^P=3/2^+$ of the
$J^P=1/2^+$ ground state $\Xi_c ^0$. The helicity angle in the
primary decay, i.e. the angle between the $\pi^+$ and the opposite
of the boost direction in the c.m. frame both calculated in the
$\Xi_c ^+(2970)$ rest frame, proved to be insensitive to some likely
$J^P$ combinations. However, the predictions for different $J^P$'s
vary significantly for the angular distributions in the secondary
decay (see Fig.~\ref{Belle}).

The analysis shows that quantum numbers $J^P=1/2^+$ are preferred
for $\Xi_c ^+(2970)$. These are the quan\-tum numbers of the Roper
resonance. The BELLE collaboration noted that its mass difference to
the $\Xi_c$ ground state is about 500 MeV. The same excitation
energy is required to excite the Roper resonance $N(1440)$, the
$\Lambda(1600)$ and the $\Sigma(1660)$, all with $J^P=1/2^+$.
\subsection{First observation of excited $\Omega_b$ states by LHCb.}
\begin{figure}[b]
\includegraphics[width=1\linewidth]{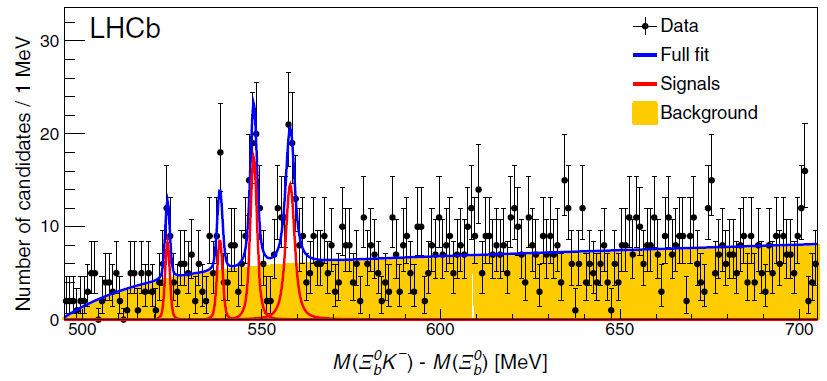}
\caption{\label{LHCb}Distribution of the mass difference
$M_{\Xi_b^0\,K^-}-M_{\Xi_b^0}$ for $\Xi_b^0\,K^-$ candidates. The
background is given by the wrong-sign candidates $\Xi_b^0\,K^+$.
(From Ref.~\cite{LHCb:2020tqd}.)\vspace{-3mm}}
\end{figure}
The LHCb collaboration searched for narrow resonances in the
$\Xi_b^0\,K^-$ invariant mass distribution~\cite{LHCb:2020tqd}. The
$\Xi_b^0$ has a lifetime of (1.48\er0.03)$10^{-12}$\,s,
$c\tau\approx 500\,\mu$m, which is sufficiently long to separate the
interaction and the decay vertices. Four peaks can be seen (Fig.~\ref{LHCb}), which
correspond to excited states of $\Omega_c$. With the given
statistics, quantum numbers can not yet be determined.

\begin{figure}[pb]
\includegraphics[width=1\linewidth]{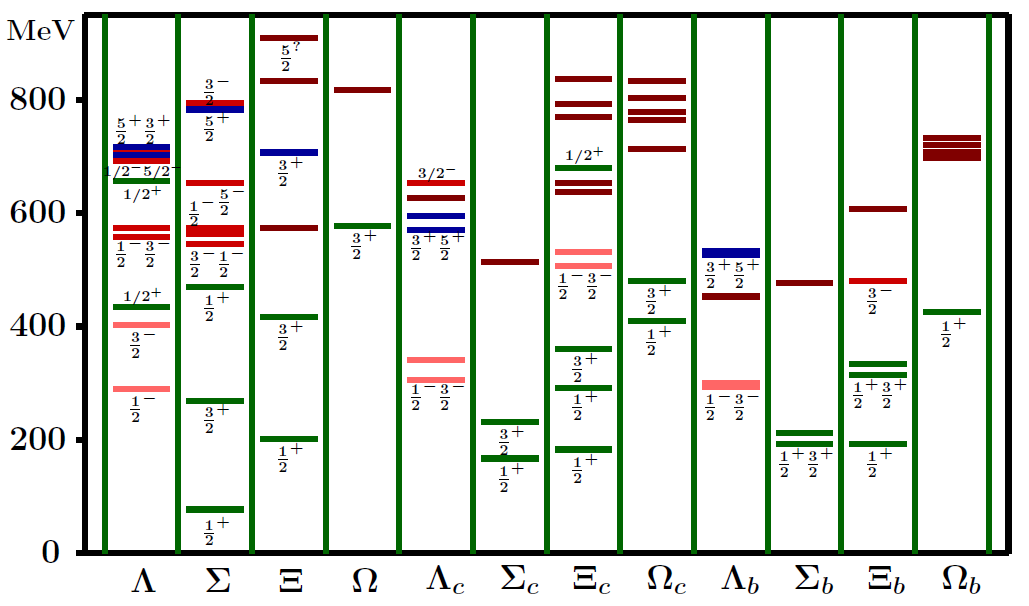}
\caption{\label{Data} Heavy baryons with charm or bottomnes and a
comparison with light baryons with strangeness. All heavy baryons
are shown, light baryons are shown at the pole mass and are only
included with 3* or 4* rating. When two quantum numbers are given,
the first one refers to the lower-mass state, the second one to the
state above. The states with $L=l_\rho=l_\lambda=0$ are shown in
green, states with $L=1$ in red (orange for members of $\bar 4_{\rm
F}$), states $L=2$ in blue, states with unknown spin-parity in
brown. }
\end{figure}
\section{The mass spectrum of excited heavy baryons}
Figure~\ref{Data} shows the mass spectrum of heavy baryons with a
single charm or bottom quark. Established light baryons with
strangeness are shown for comparison. The quantum numbers of
low-mass heavy baryons are mostly known, for higher-mass states this
information is often missing. The masses are given as excitation
energies above the $\Lambda$ ($\Lambda_c$, $\Lambda_b$) mass.

At the first glance, the spectrum looks confusing. The $\Lambda$
spectrum is crowded, there is a low-mass nega\-tive-parity spin
doublet, a second doublet -- at about the same mass  as a $\Sigma$
spin doublet -- a pair with $J^P=1/2^-$ and $5/2^-$ where a $3/2^-$
state seems to be missing, and then a positive-parity doublet with
$J^P=3/2^+, 5/2^+$. In the $\Lambda_c$ spectrum, the higher-mass
negative-parity states and the positive-parity doublet are inverted
in mass\footnote{This inversion was predicted by Capstick and Isgur
long before the states were discovered~\cite{Capstick:1986ter}}. The
$3/2^+-1/2^+$ hyperfine splitting decreases rapidly when going from
$\Sigma$ and $\Xi$ to $\Sigma_c$ and $\Xi_c$ and from $\Sigma_b$ and
$\Xi_b$. It is interesting to note that a similar pattern is
observed in mesons: the hyperfine splitting decreases when going
from $\rho-\pi$ to $D^*-D$ and to $B^*-B$. Also, there is one $\Xi$
$1/2^+$ ground state but two states for $\Xi_c$ and $\Xi_b$. The
lowest-mass $\Omega$ has $J^P=3/2^+$, in the charm sector, two
low-mass $\Omega_c$ states are known with $J^P=1/2^+$ and $3/2^+$,
the $\Omega_b$ spectrum has just one low-mass state with
$J^P=1/2^+$.

\section{\label{HBaTQS}Heavy baryons as three-quark systems}

\subsection{The spatial wave function}
The orbital wave functions of excited states are classified into two
kinds of orbital excitations, the $\lambda$-mode and the $\rho$-mode. 
In heavy baryons with one heavy quark,
the $\lambda$-mode is the excitation of the coordinate between the
heavy quark and the light diquark, and the $\rho$-mode is the
excitation of the diquark cluster. In light-baryon excitations, the
$\lambda$ and $\rho$ oscillators are mostly both excited, e.g. to
$l_\lambda=1$, $l_\rho=0$ and $l_\lambda=0$, $l_\rho=1$, the two
components of the wave function having a relative $+$ or $-$ sign.
In heavy baryons with one heavy quark, the mixing between these two
configurations is small.

The two oscillators have different reduced masses, $m_{\rho}$ and
$m_{\lambda}$:
\begin{equation}
m_{\rho}=\frac{m_q}{2}\;\;,\;\;
m_{\lambda}=\frac{2m_qm_Q}{2m_q+m_Q}.
\end{equation}
The ratio of harmonic oscillator frequencies is then given by
\begin{eqnarray}
\frac{\omega_{\lambda}}{\omega_{\rho}}=\sqrt{\frac{1}{3}(1+2m_q/m_Q)}
\leq 1 .
\label{harmo1}
\end{eqnarray}
In the heavy-quark limit ($m_Q\to\infty$), the excitation energies
in the $\lambda$ oscillator are reduced by a factor $\sqrt 3$.

\subsection{Diquarks}
We first consider the light diquark. The two light quarks can have
either the symmetric flavor structure $\mathbf{6}_F$ or the
antisymmetric flavor structure $\mathbf{\bar 3}_F$. The spin of the
light diquark can be $s_{qq} = s_l = 1$ or $s_l = 0$ leading to a
symmetric or an antisymmetric spin wave function. The color part of
the wave function is totally antisymmetric. Hence flavor and spin
wave functions are linked. In an $S$-wave, scalar (``good'' or g)
and axial-vector (``bad'' or b) diquarks can be formed. The
intrinsic quark spins couple to the internal orbital angular
momentum $l_\rho$, leading to excited diquarks with orbital
excitations.
\begin{eqnarray*}
\nonumber (l_\rho = 0,~{\bf S}) &\left\{
\begin{array}{l} s_l = 0~({\bf A}) \, , \,  \mathbf{\bar 3}_F~({\bf A}) \, , \, j_{qq} = 0 \, ,\qquad\qquad\ \hfill(\mbox{g}) \\
                 s_l = 1~({\bf S})  \, , \, \, \mathbf{6}_F~({\bf S})  \, , \, \ j_{qq} = 1 \, , \qquad \hfill(\mbox{b})
\end{array}\right. \\
         (l_\rho = 1,~{\bf A}) &\left\{
\begin{array}{l} s_l = 0~({\bf A})  \, , \, \mathbf{6}_F~({\bf S}) \, , \, \ \,j_{qq} = 1 \, , \qquad\qquad\hfill(\mbox{g})\\
                 s_l = 1~({\bf S})  \, , \, \, \mathbf{\bar 3}_F~({\bf A}) \, , \, \ j_{qq}
= 0/1/2 \, , \hfill(\mbox{b})
\end{array}\right. \\
\nonumber (l_\rho = 2,~{\bf S}) &\left\{
\begin{array}{l} s_l = 0~({\bf A}) \, , \,   \mathbf{\bar 3}_F~({\bf A}) \, , \, \ j_{qq} = 2 \, , \qquad\qquad\hfill(\mbox{g}) \\
s_l = 1~({\bf S}) \, , \, \, \mathbf{6}_F~({\bf S}) \, , \, \
\,j_{qq} = 1/2/3 \, , \hfill(\mbox{b})
\end{array}\right.
\\ \nonumber \cdots
\end{eqnarray*}
where we have denoted the total angular momentum of the light
diquark as $j_{qq}$.

\subsection{Coupling of angular momenta}
\begin{figure}[tt]
\bc
\includegraphics[width=1\linewidth]{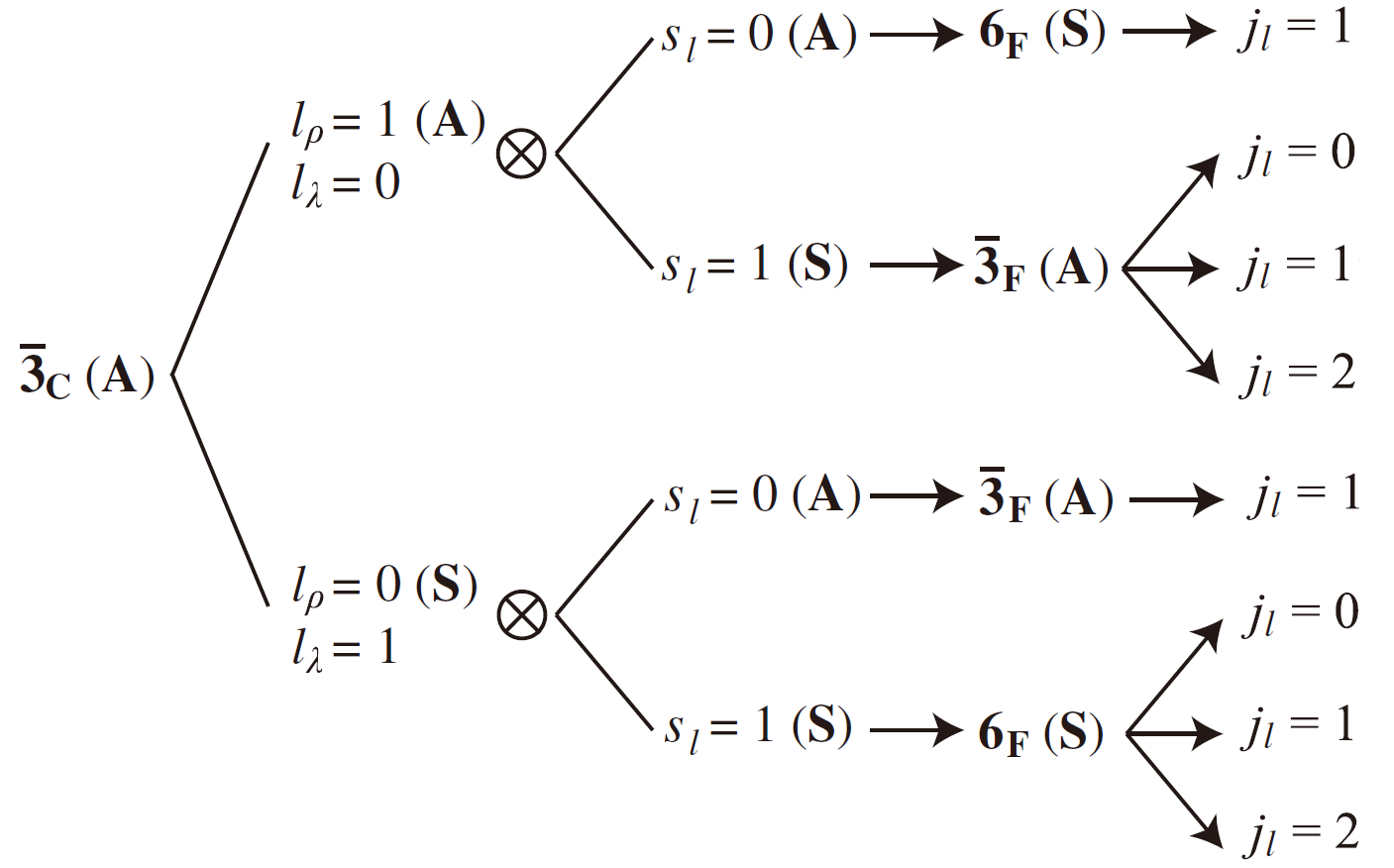}
\ec \caption{\label{Heavy-L1}Heavy baryons in $P$-wave. The light
diquark couples to the spin of the heavy quark. The light diquark of
$\Lambda$ and $\Xi$ heavy baryons are in the antisymmetric flavor
$\bar 3_{\rm F}$ and in the symmetric $6_{\rm F}$ in the case of
$\Sigma, \Xi'$ and $\Omega$ (Adapted from~\cite{Chen:2016spr}). }
\end{figure}

Figure~\ref{Heavy-L1} shows how the orbital angular momentum and the
diquark spin couple to the total diquark angular momentum $j_l$.
This in turn couples to the heavy-quark spin $s_Q$ giving rise to
spin doublets (or just spin-1/2 states for $j_l=0$). Note the
$\Lambda$ and $\Xi$ spin doublet with $s_l=0$ and $\bar 3_{\rm F}$.
In this case the wave function is antisymmetric in spin and flavor,
this is a ``good" diquark.

Only a few heavy baryons are known with $L=2$: $\Lambda_c$ and
$\Lambda_b$ with spin-parity $3/2^+$ and $5/2^+$. The other expected
states seem to show up only in the higher-mass, less-explored
region. The two observed doublets can be assigned to a configuration
in which $l_\rho=2$, $L_\lambda=0$, and the diquark is in $\bar
3_{\rm F}$ and $s_l=0$. \bc
\includegraphics[width=0.65\linewidth]{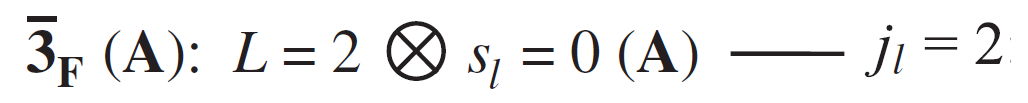}
\ec The diquark is a ``good" diquark. Note that the states
$\Lambda_c(2860)$, $\Lambda_c(2880)$ with spin-parity $3/2^+$ and
$5/2^+$ ($L=2$) are below $\Lambda_c(2940)$ with $3/2^-$. The latter
state has a ``bad" diquark and is excited to $L=1$ in the $\rho$
oscillator. In this competition, the ``good" diquark and $\lambda$
excitation with $L=2$ wins over ``bad" diquark and $\rho$ oscillator
even though the orbital angular momentum of $\Lambda_c(2940)$ is
$L=1$!

\begin{table}[ttb]
\caption{The $\lambda$- and $\rho$-mode assignments of the $P$ and
$D$-wave excitations of singly-heavy baryons. $l_\rho, l_\lambda$
are orbital angular momenta of the two oscillators, $L$ the total
orbital angular momentum, $s_q$ is the spin, $j_q$ the total angular
momentum of the diquark, and $J$ the total spin. } \label{Lr}
\centering{\footnotesize 
\begin{tabular}{cccccccc}
\hline\hline $l_\rho$&$l_\lambda$&$L$&$s_q$&$j_q$&$\Lambda,
\Xi$&\hspace{-2mm}$\Sigma,\Xi',\Omega$\hspace{-8mm}&$J^P$\\ \hline
0 & 1 & 1 & 0 & 1 &$\lambda$&$\rho$&$1/2^-, 3/2^-$\\
0 & 1 & 1 & 1 & 0 &$\rho$&$\lambda$&$1/2^-$\\
1&0& 1 & 1 & 1&$\rho$&$\lambda$&$1/2^-, 3/2^-$ \\
1&0& 1 & 1 & 2 &$\rho$&$\lambda$&$3/2^-, 5/2^-$\\ \hline
0 & 2 & 2 & 0 & 2 &$\lambda$&-&$3/2^+, 5/2^+$\\
2 & 0 & 2 & 0 & 2 &$\rho$&-&$3/2^+, 5/2^+$\\
0 & 2 & 2 & 1 & 1 &-&$\lambda$&$1/2^+, 3/2^+$\\
0&2& 2 & 1 & 2&-&$\lambda$&$3/2^+, 5/2^+$ \\
0&2& 2 & 1 & 3 &-&$\lambda$&$5/2^+, 7/2^+$\\
2 & 2 & 2 & 1 & 1 &-&$\rho$&$1/2^+, 3/2^+$\\
2&0& 2 & 1 & 2&-&$\rho$&$3/2^+, 5/2^+$ \\
2&0& 2 & 1 & 3 &-&$\rho$&$5/2^+, 7/2^+$\\ \hline\hline
\end{tabular}}
\end{table}
Table~\ref{Lr} gives a survey of the coupling scheme of $Qqq$
baryons. The spin and orbital angular momentum of the two light
quarks couple to $j_q$, and when combined with the heavy-quark spin
$s_Q$, the final $J^P$ results. There are also states with mixed
excitations like $l_\rho=1$, $l_\lambda=1$. These are unlikely to be
produced and are not included here.
$\Lambda$ and $\Xi$ with $s_q=0$ and $l_\rho=0$ have a ``good" light
diquark. For the $\Lambda_c$ we denote the light diquark by $[u,d]$.
Note that also one light and the heavy quark can be antisymmetric in
their spin and flavor wave function. We write $\Sigma_b=[ub]s$.

\subsection{Heavy quark limit}
When $m_Q\to\infty$, the heavy quark spin $s_Q$ is conserved. Due to
the conservation of the total angular momentum $J$, also the angular
momentum carried by the light quarks is conserved. Hence all
interactions which depend on the spin of the heavy quark disappear.
Thus, the mass difference within a spin doublet with, e.g.,
$J^P=3/2^+$ and $1/2^+$, will disappear in the heavy-quark limit.
Indeed, the mass differences
\begin{eqnarray}
\qquad M_{\Sigma(1520)3/2^+}&-&M_{\Sigma(1190)}=230\,{\rm MeV}\nonumber\\
\qquad M_{\Sigma_c(2520)3/2^+}&-&M_{\Sigma_c(2455)}=\ 65\,{\rm MeV}\nonumber \\
\qquad M_{\Sigma_b(5830)3/2^-}&-&M_{\Sigma_b(5820)}=\  20\,{\rm
MeV}\nonumber
\end{eqnarray}
decrease as $m_Q$ becomes large.

\section{A guide to the literature}

The first prediction of the full spectrum of baryons including
charmed and bottom baryons was presented by Capstick and
Isgur~\cite{Capstick:1986ter}, three years before the first baryon
with bottomness was discovered. The publication remained a guideline
for experimenters for now 36 years! Capstick and Isgur used a
relativized quark model with a confining potential and effective
one-gluon exchange. Based on the quark model, further studies of the
mass spectra of heavy baryons were performed. They are numerous, and
only a selction of papers can be mentioned here.

Ebert, Faustov and Galkin calculated the mass spectra for orbital
and radial excitations and constructed Regge trajectories
\cite{Ebert:2011kk}. Yu, Li, Wang, Lu, and Ya \cite{Yu:2022ymb}
calculated the mass spectra and decays of heavy baryons excited in
the $\lambda$-mode. Li, Yu, Wang, Lu, and Gu \cite{Li:2022xtj}
restricted the calculation - again based on the relativized quark
model - to the $\Xi_c$ and $\Xi_b$ families. In their model, all
excitations are in the $\lambda$-mode.

Migura, Merten, Metsch, and Petry \cite{Migura:2006en} calculated
excitations of  charmed baryons within a relativistically covariant
quark model based on the Bethe-Salpeter-equa\-tion in instantaneous
approximation. Interactions are given by a linearly rising
three-body confinement potential and a flavor dependent two-body
force derived from QCD instanton effects. Valcarce, Garcilazo and
Vijande \cite{Valcarce:2014fma} performed a comparative Faddeev
study of heavy baryons with nonrelativistic and relativistic
kinematics and different interacting potentials that differ in the
description of the hyperfine splitting. The authors conclude that
the mass difference between members of the same SU$_{\rm F}$(3)
configuration, either $\bar 3_F$  or $6_F$, is determined by the
interaction in the light-heavy quark subsystem, and the mass
difference between members of different representations is mainly
determined by the dynamics of the light diquark.

Chen, Wei and Zhang~\cite{Chen:2014nyo} derive a mass formula in a
relativistic flux tube model to calculate mass spectra for $\Lambda$
and $\Xi$ heavy baryons and assign quantum numbers to states whose
quantum numbers were not known. Faustov and Galkin
\cite{Faustov:2020gun} assigned flavor- and symmetry dependent
masses and form factors to diquarks and calculated the masses of
heavy baryons within a relativistic quark-diquark picture. Quantum
numbers are suggested for the $\Omega_c$
excitations~\cite{LHCb:2017uwr,Belle:2017ext} and other states with
unknown spin-parities. A further diquark model, again with adjusted
diquark masses, is presented by Kim, Liu, Oka, and Suzuki
\cite{Kim:2021ywp} exploiting a chiral effective theory of scalar
and vector diquarks according to the linear sigma model.

QCD sum rules have been exploited to study P-wave heavy baryons and
their decays within the heavy quark effective theory (see
\cite{Yang:2019cvw} and refs. therein). The low-lying spectrum of
charmed baryons has also been calculated in lattice QCD with a pion
mass of 156\,MeV~\cite{Bahtiyar:2020uuj}. The results - comparing
favorably with the data - are compared to earlier lattice studies
that are not discussed here.

All calculations reproduce the observed spectrum with good success,
with a large number of parameters. For the reader, it is often not
easily seen what are the main driving forces that generate the mass
spectrum. Clearly, a confinement potential is mandatory, spin
dependent forces are necessary. In the following phenomenological
part we try to identify the leading effects driving the resonance
spectrum.

\section{Phenomenology of heavy baryons}

We start with a simple observation: masses of baryons increase when
a $u$ or $d$ quark is replaced by an $s$ quark (see
Table~\ref{tab:uspin}). For light baryons, this is known as $U$-spin
rule. The constituent $s$-quark mass decreases in heavy baryons.
Note that the difference of current quark masses is $m_s-m_n\approx
124$\,MeV.
\begin{table}[t]
\caption{\label{tab:uspin}Increase of baryon masses with the number
of strange quarks. } \centering \scriptsize
\renewcommand{\arraystretch}{1.4}
\begin{tabular}{lccc}
\hline\hline
                   &$n\to s$ & $2n\to 2s$&{  $3n\to 3s$}\\\hline
$\Delta^-(1232)3/2^+$&\hspace{-1mm}$\Sigma^-(1385)3/2^+$&\hspace{-1mm}$\Xi^-(1530)3/2^+$&$ \Omega^- 3/2^+$\\
                                & +155\,MeV & +148\,MeV &  + 137\,MeV\\\hline
$\Sigma_c^0(2520)3/2^+$&\hspace{-1mm}$\Xi_c^0(2645)3/2^+$&\hspace{-1mm}$ \Omega_c^0(2770)3/2^+$&\\
                  & +128\,MeV & + 120\,MeV&\\\hline
$\Sigma_c^0(2455)1/2^+$&\hspace{-1mm}$\Xi'^0_c\ 1/2^+$&\hspace{-1mm}$\Omega_c^0\ 1/2^+$&\\
                     & +121\,MeV& +116\,MeV&\\\hline
$\Sigma_b^-(5816)1/2^+$&\hspace{-1mm}$\Xi'^0_b\ 1/2^+$&\hspace{-1mm}$\Omega_b^- 1/2^+$&\\
                     & +120\,MeV& +111\,MeV&\\\hline\hline
\end{tabular}
\end{table}

In Table~\ref{tab:HF} we show the mass difference of the lowest-mass
$J^P=3/2^-$ states with ($u,d,s,c$) or ($u,d,s,b$) quarks and the
$J^P=1/2^+$ ground states: The mass differences are surprisingly
small. The $N(1520)-N$ mass difference is 580 MeV, much larger than
the mass differences seen here. In the table, [ud] represents wave
functions with a $u,d$ quark pair that is anti-symmetric  in spin
and flavor. These diquarks are often called {\it good diquarks}. The
presence of good diquarks leads to a stronger binding. In the
4-plet, all three quark pairs have such a component w.r.t. their
exchange. We denote this by [ud,us,ds]. Thus there are three good
diquarks in the wave function. This fact leads to the low masses of
the 4-plet members. The similarity of the mass splittings supports
similar interpretations of the four resonances from  $\Lambda(1520)$
to $\Xi^0 _{b\ 3/2^-}$.

\begin{table}[t]
\caption{\label{tab:HF}Mass splitting between baryon ground states
belonging to the symmetric 20plet (with $J^P=3/2^+$) and to the
mixed-symmetry 20plet (with $J^P=1/2^+$).} \centering \scriptsize
\renewcommand{\arraystretch}{1.4}
\begin{tabular}{ccccc}
\hline\hline
$\Xi^0 _{b\ 3/2^-}$&\hspace{-2mm}[us,ub,sb]&$\Xi^0 _{b\ 1/2^+}$&\hspace{-2mm}[us]&$\delta M = 310$\,MeV\\
$\Lambda^0 _{b\ 3/2^-}$&\hspace{-2mm}[ud,ub,db]&$\Lambda^0 _{b\ 1/2^+}$&\hspace{-2mm}[ud]&$\delta M = 300$\,MeV\\
$\Xi^+ _{c\ 3/2^-}$&\hspace{-2mm}[us,uc,sc]&$\Xi^+ _{c\ 1/2^+}$&\hspace{-2mm}[us]&$\delta M = 350$\,MeV\\
$\Lambda^+ _{c\ 3/2^-}$&\hspace{-2mm}[ud,uc,dc]&$\Lambda^+ _{c\ 1/2^+}$&\hspace{-2mm}[ud]& $\delta M = 400$\,MeV\\
$\Lambda(1520)$&\hspace{-2mm}[ud,us,ds]&$\Lambda_{1/2^+}$&\hspace{-2mm}[ud]& $\delta M = 400$\,MeV\\
\hline\hline
\end{tabular}
\end{table}

In most publications, both resonances, $\Lambda_c(2595)1/2^-$ and
$\Lambda_c(2625)3/2^-$, are discussed as 3-quark baryons. However,
Nieves and Pavao~\cite{Nieves:2019nol} have studied these two
resonances in an effective field theory that incorporates the
interplay between $\Sigma^{(*)}_c \pi - ND^{(*)}$ baryon-meson
dynamics and bare $P$-wave $cud$ quark-model state and suggest that
these two resonances are not heavy quark symmetry spin partners.
Instead, they see $\Lambda_c(2625)3/2^-$ as a dressed three-quark
state while $\Lambda_c(2595)1/2^-$ is reported to have a predominant
molecular structure. Nevertheless, the two states
$\Lambda_c(2625)3/2^-$ and $\Lambda_c(2595)1/2^-$ obviously form a
spin doublet.

The mass shift in H atoms between the two ground states with
electron and proton spins parallel or antiparallel is called
hyperfine splitting. We borrow this expression to discuss the
difference between the ground states with all three quark spins
adding to $J=3/2$ (and belonging to the symmetric 20-plet) and with
those having $J=1/2$ (that belong to the mixed-symmetrx 20-plet). We
thus compare masses of the fully symmetric $20_s$-plet with those
from the $\bar 3$-plet or 6-plet within the $20_m$-plet (see
Table~\ref{tab:symm}). The two configurations differ by the
orientation of the heavy-quark spin relative to the spin of the
light diquark. According to the heavy-quark-spin symmetry, this mass
difference has to vanish with $m_Q\to\infty$. In the Table we assume
constituent quark masses of 0.15\,GeV ($u,d$), 0.3\,GeV ($s$),
1.25\,GeV ($c$) and 4.1\,GeV ($b$).

The $J^P=3/2^+$ states have a fully symmetric flavor wave function,
the $J^P=1/2^+$ states have an antisymmetric quark pair (a good
diquark) that is indicated in the list. Their effect scales with
$1/m_q$. The mass shift due to the presence of good diquarks is
expected for instanton-induced interactions.

\begin{table}[t]
\caption{\label{tab:symm}Mass splitting between baryons with fully
symmetric wave functions and baryons with antisymmetric quark pairs.
The [us] indicates an antismmetric quark pair. }
\renewcommand{\arraystretch}{1.4}
\centering \scriptsize
\begin{tabular}{cccccc}
\hline\hline
                  &              &       & $\delta M$  &  $m_q$ & $\delta M\cdot m_q$\\\hline
$\Sigma_{b\ 3/2^+}$& $\Lambda_b$&$[ud]b$   &0.211\,MeV &$\sim 0.3$\,GeV & 0.063\\
$\Sigma_{c\ 3/2^+}$& $\Lambda_c$&$[ud]c$   &0.232\,MeV  &$\sim 0.3$\,GeV & 0.070\\
$\Sigma_{\ 3/2^+}$ & $\Lambda   $&$[ud]s$  &0.268\,MeV  &$\sim 0.3$\,GeV& 0.080\\
$\Delta_{\ 3/2^+}$  & N      &$[ud]u$      &0.292\,MeV  &$\sim
0.3$\,GeV & 0.088\\\hline
$\Xi_{b\ 3/2^+}$& $\Xi_b$&$[us]b$          &0.163\,MeV  &$\sim 0.45$\,GeV & 0.073\\
$\Xi_{c\ 3/2^+}$      & $\Xi_c $&$[us]c$    &0.177\,MeV  &$\sim 0.45$\,GeV & 0.080\\
$\Xi_{\ 3/2^+}$      & $\Xi $&$[us]s$       &0.217\,MeV  &$\sim 0.45$\,GeV & 0.098\\
$\Sigma_{\ 3/2^+}$ & $\Sigma    $&$[us]$u  &0.191\,MeV  &$\sim
0.45$\,GeV& 0.086\\\hline
$\Xi_{c\ 3/2^+}$      & $\Xi_c ^{'}$&$[uc]s$&0.067\,MeV  &$\sim 1.4$\,GeV & 0.093\\
$\Sigma_{c\ 3/2^+}$& $\Sigma_c$&$[uc]u$    &0.065\,MeV  &$\sim
1.4$\,GeV &0.090\\\hline
$\Xi_{b\ 3/2^+}$& $\Xi_b ^{'}$&$[ub]s$     &0.020\,MeV  &$\sim 4.25$\,GeV &0.085\\
$\Sigma_{b\ 3/2^+}$& $\Sigma_b$&$[ub]u$    &0.021\,MeV  &$\sim 4.25$\,GeV &0.089\\
\hline\hline
\end{tabular}
\end{table}
\subsection{Heavy baryons at higher mass:}
Next we discuss the higher-mass negative-parity states. In
light-baryon spectroscopy, there are seven negative-parity $\Lambda$
states expected in the first excitation level: two singlet states
with $J^P=1/2^-,3/2^-$, two octet states with intrinsic total quark
spin $s=1/2$ and $J^P=1/2^-,3/2^-$, and a $J^P=1/2^-,3/2^-,5/2^-$
triplet with $s=3/2$. In light baryons, both $\lambda$ and $\rho$
oscillator are coherently excited. In heavy-quark baryons, the two
oscillators decouple, and the $\lambda$ and $\rho$ modes are well
separated. The low-lying spin-doublet of $P$-wave $\Lambda_Q$ states
is dominated by a $\lambda$-mode excitation, the other five expected
states are excited in the $\rho$ mode.

Unfortunately, only one negative-parity state at a higher mass has
been reported, the $\Lambda_c(2940)3/2^-$. Its mass is 653 MeV above
the $\Lambda_c ^+$.  We interpret this state as $l_\rho$ excitation
with a diquark spin $s=1$. The $\Lambda(1690)3/2^-$ is only 570\,MeV
above the $\Lambda$, it is excited in both the $\lambda$ and the
$\rho$ mode.

The mass of $\Lambda_c(2940)3/2^-$ (with intrinsic orbital angular
momentum $L=1$) is above the masses of the positive-parity states
$\Lambda_c(2860)3/2^+$ and $\Lambda_c(2880)5/2^+$ (having $L=2$).
Yet, the mass of $\Lambda(1690)3/2^-$ falls well below the masses of
$\Lambda(1890)3/2^+$ and $\Lambda(1820)5/2^+$ for reasons discussed
above.

\section{Pentaquarks}
\begin{figure}[t]
\bc
\includegraphics[width=0.4\textwidth]{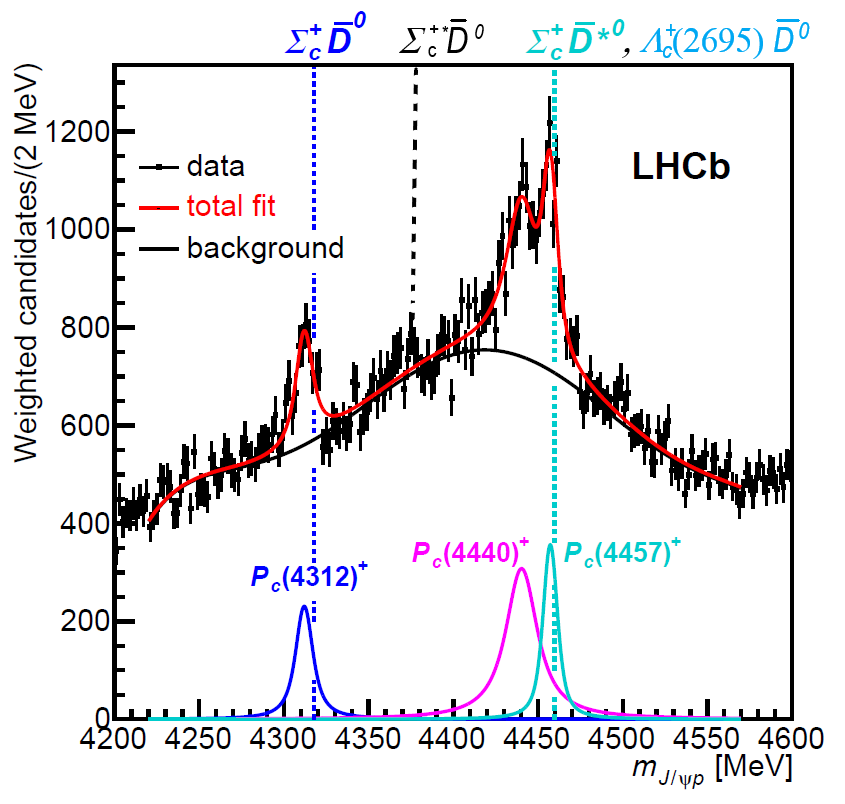}
\vspace{-2mm} \ec \caption{\label{Penta1}(Color online) The
$J/\psi\,p$ mass distribution fitted with three BW amplitudes and a
sixth-order polynomial background. The thresholds for the $\Sigma_c
^+\bar D ^0$. (Adapted from \cite{LHCb:2019kea}.)}
\end{figure}
In 2015, the LHCb collaboration reported the observation of two
exotic structures in the $J/\psi p$ system, a broad  resonant
structure with a Breit-Wigner width of about $200\,\mbox{MeV}$
 called $P_c(4380)^+$ and  a narrow state, $P_c(4450)^+$~\cite{LHCb:2015yax}. The exotic structures
 were observed in the reaction $\Lambda_b ^0\to J/\psi K^- p$.
An excited three-quark nucleon cannot decay into $J/\psi\,p$, this
would violate the OZI rule. Hence the minimal quark content is
($c\bar cuud$). The findings met with great interest; the
publication is quoted nearly 1500 times (2022, November). Indeed
narrow baryonic resonances with hidden charm had been predicted
several years before as dynamically generated states
\cite{Hofmann:2006qx,Wu:2010vk,Wu:2012md}.

A multitude of different interpretations of the observed structures
is offered in the literature, but none is accepted anonymously.
There are numerous reviews on tetra- and pentaquarks and their
possible interpretations
\cite{Chen:2016qju,Chen:2016qju,Guo:2017jvc,Olsen:2017bmm,Brambilla:2019esw,Liu:2019zoy,Burns:2022uiv}.

With increased statistics, $P_c(4312)^+$ was confirmed and the
higher-mass $P_c(4450)^+$ was shown to be split into two narrow
overlapping structures, $P_c(4440)^+$ and $P_c(4457)^+$
\cite{LHCb:2019kea}. The existence of the broad resonance was not
confirmed. The data and a fit are shown in Fig.~\ref{Penta1} which
also displays some relevant thresholds. In addition, a further
smaller structure can be seen at 4380\,MeV, close to the
$\Sigma_c^{+*}\bar{D}^0$ threshold. A narrow structure here is
expected in molecular models (see e.~g.~ \cite{Du:2021fmf}), but due
to limited statistics there was no attempt to describe it in the
recent LHCb analysis \cite{LHCb:2019kea}. The resonant parameters --
including the broad structure at 4380\,MeV -- are reproduced in
Table~\ref{tab:Penta}.

\begin{table}[t]
\caption{\label{tab:Penta}$J/\psi$ resonances found by the LHCb
collaboration.}
\renewcommand{\arraystretch}{1.2}
\centering
\begin{tabular}{lcrl}
\hline\hline
$P_c(4312)^+$:&M =& (4311.9 \er 0.7 $^{+6.8}_{-0.6}$) &\mbox{ MeV} \\
\cite{LHCb:2019kea}  &$\Gamma$ =&  (9.8 \er 2.7 $^{+3.7}_{-4.5}$) &\mbox{ MeV}\\
$P_c(4380)^+$:& M = & (4380 \er 30) &\mbox{ MeV} \\
\cite{LHCb:2015yax} &$\Gamma$ =&  (205 \er 90) &\mbox{ MeV} \\
$P_c(4440)^+$:& M =& (4440.3 \er 1.3 $^{+4.1}_{-4.7}$) &\mbox{ MeV} \\
\cite{LHCb:2019kea} &$\Gamma$ =& (20.6 \er 4.9 $^{+8.7}_{-10.1}$) &\mbox{ MeV} \\
$P_c(4457)^+$:& M =& (4457.3 \er 0.6 $^{+4.1}_{-1.7}$) &\mbox{ MeV} \\
\cite{LHCb:2019kea} &$\Gamma$ =&  (6.4 \er 2.0 $^{+5.7}_{-1.9}$) &\mbox{ MeV} \\
$P_c(4337)^+$:&M =& (4337 $^{+7}_{-4}{}^{+2}_{-2}$)&\mbox{MeV}\\
\cite{LHCb:2021chn}&$\Gamma$ =&(29 $^{+26}_{-12}{}^{+14}_{-14}$)&\mbox{ MeV} \\
 \hline\hline\\
\end{tabular}
\end{table}

Quantum numbers $J^P=3/2^-$ and $5/2^+$ were preferred for
$P_c(4380)^+$ and $P_c(4450)^+$. In the later publication
\cite{LHCb:2019kea}, no quantum numbers are determined.

In the reaction $B_s ^0\to J/\psi\,\bar p\,p$ a pentaquark-like
structure, named $P_c(4337)^+$, was observed in the $J/\psi\,\bar p$
and  $J/\psi\,p$ mass distributions \cite{LHCb:2021chn}. The
significance, as determined from a 3-body amplitude analysis, is
between 3.1 and 3.7$\,\sigma$. Its Breit-Wigner parameters are
incompatible with the structures observed in $\Lambda_b$ decays. The
lighter state at $4312\,\mbox{MeV}$ was not found in this reaction,
highlighting the importance of the production mechanism for the
formation of these resonances.  However, it has been pointed out in
\cite{Wang:2021crr} that in a region with many close-by thresholds,
the Breit-Wigner parameters measured in a particular channel may
differ significantly from the pole location.

Strange counterparts to these pentaquark states have been searched
for by LHCb in the reaction $\Xi_b^-\to J/\psi\,\Lambda
K^-$ \cite{LHCb:2020jpq}. Resonances in the $J/\psi\,\Lambda$ system
are denoted by $P_{cs} ^0(4312)$ and have ($c\bar c uds$) as minimal
quark content. A peak was found close to the $\Xi_c ^0\,D^{*0}$
threshold with a mass and width given in Table~\ref{tab:penta2}.

The $J/\psi \Lambda$ system was also investigated in 2019 by CMS
\cite{CMS:2019kbn}, exploiting the small phase space available in
the B-meson decay  $B^- \to J/\psi\,\Lambda \bar p$. The analysis
showed that the observed spectrum was incompatible with a pure phase
space distribution. Very recently, the LHCb collaboration reported a
new analysis of this process~\cite{LHCb:2022jad}. Now, a signal in
the $J/\psi\,\Lambda$ subsystem, with preferred quantum numbers
$J^P=1/2^-$, was established at high significance, named
$P_{cs}^0(4338)$. Due to the presence of the second (anti)baryon,
the phase space in the $B$-meson decay is too small to access the
heavier pentaquark state found in the $\Xi_b$ decay.

\begin{table}[t]
\caption{\label{tab:penta2}$J/\psi\Lambda$ pentaquarks. }
\renewcommand{\arraystretch}{1.2}
\centering
\begin{tabular}{lcrl}
\hline\hline
\hspace{-3mm}$P_{cs}^0(4459)$:&M =        &(4458.8 \er 2.9 $^{+4.7}_{-1.1}$) &\mbox{MeV}\\
\cite{LHCb:2020jpq}           &$\Gamma$ = &(17.3 \er 6.5 $^{+8.0}_{-5.7}$) &\mbox{MeV}\\
\hspace{-3mm}$P_{cs}^0(4338)$:&M =         &(4338.2 \er 0.7 \er 0.4) &\mbox{MeV}\\
\cite{LHCb:2022tbp}           &$\Gamma$ = &(7.0 \er 1.2 \er 1.3) &\mbox{MeV}\\
\hline\hline\\
\end{tabular}
\end{table}

These structures have stimulated an intense discussion of the nature
of these structures. Do they originate from threshold singularities
due to rescattering in the final state leading to a logarithmic
branching point in the amplitude? Are they hadronic molecules like
the deuteron? Are they compact or triple-quark--diquark systems or
states where a $c\bar c$ center is surrounded by light quarks?

The peaks are mostly seen very close to important thresholds. Thus
they could originate from threshold singularities. We refer to a few
publications~\cite{Guo:2015umn,Liu:2015fea,Bayar:2016ftu,Ali:2017jda}.
The LHCb collaboration studied this hypothesis and found it
incompatible with the data, but the attempts continued
\cite{Guo:2019twa,Dong:2020hxe,Shen:2020gpw,Nakamura:2021qvy}.

Very popular are interpretations as bound states composed of charmed
baryons and anti-charmed mesons or of charmonium states binding
light-quark baryons. The pentaquark states are then seen to be of
molecular nature and be bound by coupled-channel dynamics
\cite{Eides:2015dtr,Guo:2019fdo,Du:2019pij,Xu:2020gjl,Chen:2020kco,Chen:2020uif,Wu:2021caw,Lu:2021irg,Du:2021fmf,Zhu:2021lhd,Malabarba:2021taj,Yalikun:2021bfm}.
Diquark-triquark models were studied
\cite{Zhu:2015bba,Ali:2019npk,Shi:2021wyt,Azizi:2021utt}, and sum
rules are exploited in Refs.~\cite{Wang:2019got,Wang:2020eep}.
\section{Concluding remarks:}
The study of hadrons with heavy quarks has developed into a
fascinating new field of particle physics. Particular excitement is
due to the discovery of unconventional structures that are hotly
debated. But also the ``regular" heavy hadrons yield very useful
information on the interactions of quarks in the confinement region.

\bibliography{heavies}

\begin{thebibliography}{10}

\bibitem{ParticleDataGroup:2022pth}
R.~L. Workman {\em et~al.}, ``{Review of Particle Physics},'' {\em PTEP},
  vol.~2022, p.~083C01, 2022.

\bibitem{LHCb:2015yax}
R.~Aaij {\em et~al.}, ``{Observation of $J/\psi p$ Resonances Consistent with
  Pentaquark States in $\Lambda_b^0 \to J/\psi K^- p$ Decays},'' {\em Phys.
  Rev. Lett.}, vol.~115, p.~072001, 2015.

\bibitem{LHCb:2019kea}
R.~Aaij {\em et~al.}, ``{Observation of a narrow pentaquark state,
  $P_c(4312)^+$, and of two-peak structure of the $P_c(4450)^+$},'' {\em Phys.
  Rev. Lett.}, vol.~122, no.~22, p.~222001, 2019.

\bibitem{Gratrex:2022xpm}
J.~Gratrex, B.~Meli\'c, and I.~Ni\v{s}and\v{z}i\'c, ``{Lifetimes of singly
  charmed hadrons},'' 2022.

\bibitem{SELEX:2002wqn}
M.~Mattson {\em et~al.}, ``{First Observation of the Doubly Charmed Baryon
  $\Xi^+_{cc}$},'' {\em Phys. Rev. Lett.}, vol.~89, p.~112001, 2002.

\bibitem{SELEX:2004lln}
A.~Ocherashvili {\em et~al.}, ``{Confirmation of the double charm baryon
  $\Xi^+(cc)(3520)$ via its decay to $p D^+ K^-$},'' {\em Phys. Lett. B},
  vol.~628, pp.~18--24, 2005.

\bibitem{LHCb:2019epo}
R.~Aaij {\em et~al.}, ``{Precision measurement of the $\Xi_{cc}^{++}$ mass},''
  {\em JHEP}, vol.~02, p.~049, 2020.

\bibitem{LHCb:2021eaf}
R.~Aaij {\em et~al.}, ``{Search for the doubly charmed baryon $
  {\varXi}_{cc}^{+} $ in the $ {\varXi}_c^{+}{\pi}^{-}{\pi}^{+} $ final
  state},'' {\em JHEP}, vol.~12, p.~107, 2021.

\bibitem{LHCb:2022fbu}
R.~Aaij {\em et~al.}, ``{Search for the doubly heavy baryon $\it{\Xi}_{bc}^{+}$
  decaying to $J/\it{\psi} \it{\Xi}_{c}^{+}$},'' 4 2022.

\bibitem{BaBar:2006pve}
B.~Aubert {\em et~al.}, ``{Observation of an excited charm baryon $\Omega_c ^*$
  decaying to $\Omega_c ^0\gamma$},'' {\em Phys. Rev. Lett.}, vol.~97,
  p.~232001, 2006.

\bibitem{Belle:2020tom}
T.~J. Moon {\em et~al.}, ``{First determination of the spin and parity of the
  charmed-strange baryon $\Xi_{c}(2970)^+$},'' {\em Phys. Rev. D}, vol.~103,
  no.~11, p.~L111101, 2021.

\bibitem{LHCb:2020tqd}
R.~Aaij {\em et~al.}, ``{First observation of excited $\Omega_b^-$ states},''
  {\em Phys. Rev. Lett.}, vol.~124, no.~8, p.~082002, 2020.

\bibitem{Capstick:1986ter}
S.~Capstick and N.~Isgur, ``{Baryons in a relativized quark model with
  chromodynamics},'' {\em Phys. Rev. D}, vol.~34, no.~9, pp.~2809--2835, 1986.

\bibitem{Chen:2016spr}
H.-X. Chen, W.~Chen, X.~Liu, Y.-R. Liu, and S.-L. Zhu, ``{A review of the open
  charm and open bottom systems},'' {\em Rept. Prog. Phys.}, vol.~80, no.~7,
  p.~076201, 2017.

\bibitem{Ebert:2011kk}
D.~Ebert, R.~N. Faustov, and V.~O. Galkin, ``{Spectroscopy and Regge
  trajectories of heavy baryons in the relativistic quark-diquark picture},''
  {\em Phys. Rev. D}, vol.~84, p.~014025, 2011.

\bibitem{Yu:2022ymb}
G.-L. Yu, Z.-Y. Li, Z.-G. Wang, J.~Lu, and M.~Yan, ``{Systematic analysis of
  single heavy baryons $\Lambda_{Q}$, $\Sigma_{Q}$ and $\Omega_{Q}$},'' 6 2022.

\bibitem{Li:2022xtj}
Z.-Y. Li, G.-L. Yu, Z.-G. Wang, J.~Lu, and J.-Z. Gu, ``{Systematic analysis of
  strange single heavy baryons},'' 7 2022.

\bibitem{Migura:2006en}
S.~Migura, D.~Merten, B.~Metsch, and H.-R. Petry, ``{Semileptonic decays of
  baryons in a relativistic quark model},'' {\em Eur. Phys. J. A}, vol.~28,
  p.~55, 2006.

\bibitem{Valcarce:2014fma}
A.~Valcarce, H.~Garcilazo, and J.~Vijande, ``{Heavy baryon spectroscopy with
  relativistic kinematics},'' {\em Phys. Lett. B}, vol.~733, pp.~288--295,
  2014.

\bibitem{Chen:2014nyo}
B.~Chen, K.-W. Wei, and A.~Zhang, ``{Assignments of $\Lambda_Q$ and $\Xi_Q$
  baryons in the heavy quark-light diquark picture},'' {\em Eur. Phys. J. A},
  vol.~51, p.~82, 2015.

\bibitem{Faustov:2020gun}
R.~N. Faustov and V.~O. Galkin, ``{Heavy Baryon Spectroscopy in the
  Relativistic Quark Model},'' {\em Particles}, vol.~3, no.~1, pp.~234--244,
  2020.

\bibitem{LHCb:2017uwr}
R.~Aaij {\em et~al.}, ``{Observation of five new narrow $\Omega_c^0$ states
  decaying to $\Xi_c^+ K^-$},'' {\em Phys. Rev. Lett.}, vol.~118, no.~18,
  p.~182001, 2017.

\bibitem{Belle:2017ext}
J.~Yelton {\em et~al.}, ``{Observation of Excited $\Omega_c$ Charmed Baryons in
  $e^+e^-$ Collisions},'' {\em Phys. Rev. D}, vol.~97, no.~5, p.~051102, 2018.

\bibitem{Kim:2021ywp}
Y.~Kim, Y.-R. Liu, M.~Oka, and K.~Suzuki, ``{Heavy baryon spectrum with chiral
  multiplets of scalar and vector diquarks},'' {\em Phys. Rev. D}, vol.~104,
  no.~5, p.~054012, 2021.

\bibitem{Yang:2019cvw}
H.-M. Yang, H.-X. Chen, E.-L. Cui, A.~Hosaka, and Q.~Mao, ``{Decay properties
  of $P$-wave bottom baryons within light-cone sum rules},'' {\em Eur. Phys. J.
  C}, vol.~80, no.~2, p.~80, 2020.

\bibitem{Bahtiyar:2020uuj}
H.~Bahtiyar, K.~U. Can, G.~Erkol, P.~Gubler, M.~Oka, and T.~T. Takahashi,
  ``{Charmed baryon spectrum from lattice QCD near the physical point},'' {\em
  Phys. Rev. D}, vol.~102, no.~5, p.~054513, 2020.

\bibitem{Nieves:2019nol}
J.~Nieves and R.~Pavao, ``{Nature of the lowest-lying odd parity charmed baryon
  $\Lambda_c(2595)$ and $\Lambda_c(2625)$ resonances},'' {\em Phys. Rev. D},
  vol.~101, no.~1, p.~014018, 2020.

\bibitem{Hofmann:2006qx}
J.~Hofmann and M.~F.~M. Lutz, ``{D-wave baryon resonances with charm from
  coupled-channel dynamics},'' {\em Nucl. Phys. A}, vol.~776, pp.~17--51, 2006.

\bibitem{Wu:2010vk}
J.-J. Wu, R.~Molina, E.~Oset, and B.~S. Zou, ``{Dynamically generated $N^{*}$
  and $\Lambda^*$ resonances in the hidden charm sector around 4.3 GeV},'' {\em
  Phys. Rev. C}, vol.~84, p.~015202, 2011.

\bibitem{Wu:2012md}
J.-J. Wu, T.~S.~H. Lee, and B.~S. Zou, ``{Nucleon Resonances with Hidden Charm
  in Coupled-Channel Models},'' {\em Phys. Rev. C}, vol.~85, p.~044002, 2012.

\bibitem{Chen:2016qju}
H.-X. Chen, W.~Chen, X.~Liu, and S.-L. Zhu, ``{The hidden-charm pentaquark and
  tetraquark states},'' {\em Phys. Rept.}, vol.~639, pp.~1--121, 2016.

\bibitem{Guo:2017jvc}
F.-K. Guo, C.~Hanhart, U.-G. Mei\ss{}ner, Q.~Wang, Q.~Zhao, and B.-S. Zou,
  ``{Hadronic molecules},'' {\em Rev. Mod. Phys.}, vol.~90, no.~1, p.~015004,
  2018.
\newblock [Erratum: Rev.Mod.Phys. 94, 029901 (2022)].

\bibitem{Olsen:2017bmm}
S.~L. Olsen, T.~Skwarnicki, and D.~Zieminska, ``{Nonstandard heavy mesons and
  baryons: Experimental evidence},'' {\em Rev. Mod. Phys.}, vol.~90, no.~1,
  p.~015003, 2018.

\bibitem{Brambilla:2019esw}
N.~Brambilla, S.~Eidelman, C.~Hanhart, A.~Nefediev, C.-P. Shen, C.~E. Thomas,
  A.~Vairo, and C.-Z. Yuan, ``{The $XYZ$ states: experimental and theoretical
  status and perspectives},'' {\em Phys. Rept.}, vol.~873, pp.~1--154, 2020.

\bibitem{Liu:2019zoy}
Y.-R. Liu, H.-X. Chen, W.~Chen, X.~Liu, and S.-L. Zhu, ``{Pentaquark and
  Tetraquark states},'' {\em Prog. Part. Nucl. Phys.}, vol.~107, pp.~237--320,
  2019.

\bibitem{Burns:2022uiv}
T.~J. Burns and E.~S. Swanson, ``{Production of $P_c$ states in $\Lambda_b$
  decays},'' {\em Phys. Rev. D}, vol.~106, no.~5, p.~054029, 2022.

\bibitem{Du:2021fmf}
M.-L. Du, V.~Baru, F.-K. Guo, C.~Hanhart, U.-G. Mei\ss{}ner, J.~A. Oller, and
  Q.~Wang, ``{Revisiting the nature of the P$_{c}$ pentaquarks},'' {\em JHEP},
  vol.~08, p.~157, 2021.

\bibitem{LHCb:2021chn}
R.~Aaij {\em et~al.}, ``{Evidence for a new structure in the $J/\psi p$ and
  $J/\psi \bar{p}$ systems in $B_s^0 \to J/\psi p \bar{p}$ decays},'' {\em
  Phys. Rev. Lett.}, vol.~128, no.~6, p.~062001, 2022.

\bibitem{Wang:2021crr}
J.-Z. Wang, X.~Liu, and T.~Matsuki, ``{Evidence supporting the existence of
  $P_c(4380)^{\pm}$ from the recent measurements of $B_s\to J/\psi p\bar p$},''
  {\em Phys. Rev. D}, vol.~104, no.~11, p.~114020, 2021.

\bibitem{LHCb:2020jpq}
R.~Aaij {\em et~al.}, ``{Evidence of a $J/\psi\Lambda$ structure and
  observation of excited $\Xi^-$ states in the $\Xi^-_b \to J/\psi\Lambda K^-$
  decay},'' {\em Sci. Bull.}, vol.~66, pp.~1278--1287, 2021.

\bibitem{CMS:2019kbn}
A.~M. Sirunyan {\em et~al.}, ``{Study of the $ {\mathrm{B}}^{+}\to
  \mathrm{J}/\psi \overline{\Lambda}\mathrm{p} $ decay in proton-proton
  collisions at $ \sqrt{s} $ = 8 TeV},'' {\em JHEP}, vol.~12, p.~100, 2019.

\bibitem{LHCb:2022jad}
``{Observation of a $J/\psi\Lambda$ resonance consistent with a strange
  pentaquark candidate in $B^-\to J/\psi\Lambda\bar{p}$ decays},'' 10 2022.

\bibitem{LHCb:2022tbp}
C.~Chen and E.~S. Norella, ``{Particle Zoo 2.0: New Tetra- and Pentaquarks at
  LHCb},'' {\em CERN Seminar}, pp.~July, 5, 2022.

\bibitem{Guo:2015umn}
F.-K. Guo, U.-G. Mei\ss{}ner, W.~Wang, and Z.~Yang, ``{How to reveal the exotic
  nature of the P$_c$(4450)},'' {\em Phys. Rev. D}, vol.~92, no.~7, p.~071502,
  2015.

\bibitem{Liu:2015fea}
X.-H. Liu, Q.~Wang, and Q.~Zhao, ``{Understanding the newly observed heavy
  pentaquark candidates},'' {\em Phys. Lett. B}, vol.~757, pp.~231--236, 2016.

\bibitem{Bayar:2016ftu}
M.~Bayar, F.~Aceti, F.-K. Guo, and E.~Oset, ``{A Discussion on Triangle
  Singularities in the $\Lambda_b \to J/\psi K^{-} p$ Reaction},'' {\em Phys.
  Rev. D}, vol.~94, no.~7, p.~074039, 2016.

\bibitem{Ali:2017jda}
A.~Ali, J.~S. Lange, and S.~Stone, ``{Exotics: Heavy Pentaquarks and
  Tetraquarks},'' {\em Prog. Part. Nucl. Phys.}, vol.~97, pp.~123--198, 2017.

\bibitem{Guo:2019twa}
F.-K. Guo, X.-H. Liu, and S.~Sakai, ``{Threshold cusps and triangle
  singularities in hadronic reactions},'' {\em Prog. Part. Nucl. Phys.},
  vol.~112, p.~103757, 2020.

\bibitem{Dong:2020hxe}
X.-K. Dong, F.-K. Guo, and B.-S. Zou, ``{Explaining the Many Threshold
  Structures in the Heavy-Quark Hadron Spectrum},'' {\em Phys. Rev. Lett.},
  vol.~126, no.~15, p.~152001, 2021.

\bibitem{Shen:2020gpw}
C.-W. Shen, H.-J. Jing, F.-K. Guo, and J.-J. Wu, ``{Exploring Possible Triangle
  Singularities in the $\Xi^-_{b} \to K^- J/\psi \Lambda$ Decay},'' {\em
  Symmetry}, vol.~12, no.~10, p.~1611, 2020.

\bibitem{Nakamura:2021qvy}
S.~X. Nakamura, ``{$P_c(4312)^+$, $P_c(4380)^+$, and $P_c(4457)^+$ as double
  triangle cusps},'' {\em Phys. Rev. D}, vol.~103, p.~111503, 2021.

\bibitem{Eides:2015dtr}
M.~I. Eides, V.~Y. Petrov, and M.~V. Polyakov, ``{Narrow Nucleon-$\psi(2S)$
  Bound State and LHCb Pentaquarks},'' {\em Phys. Rev. D}, vol.~93, no.~5,
  p.~054039, 2016.

\bibitem{Guo:2019fdo}
F.-K. Guo, H.-J. Jing, U.-G. Mei\ss{}ner, and S.~Sakai, ``{Isospin breaking
  decays as a diagnosis of the hadronic molecular structure of the
  $P_c(4457)$},'' {\em Phys. Rev. D}, vol.~99, no.~9, p.~091501, 2019.

\bibitem{Du:2019pij}
M.-L. Du, V.~Baru, F.-K. Guo, C.~Hanhart, U.-G. Mei\ss{}ner, J.~A. Oller, and
  Q.~Wang, ``{Interpretation of the LHCb $P_c$ States as Hadronic Molecules and
  Hints of a Narrow $P_c(4380)$},'' {\em Phys. Rev. Lett.}, vol.~124, no.~7,
  p.~072001, 2020.

\bibitem{Xu:2020gjl}
H.~Xu, Q.~Li, C.-H. Chang, and G.-L. Wang, ``{Recently observed $P_c$ as
  molecular states and possible mixture of $P_c(4457)$},'' {\em Phys. Rev. D},
  vol.~101, no.~5, p.~054037, 2020.

\bibitem{Chen:2020kco}
R.~Chen, ``{Can the newly reported $P_{cs}(4459)$ be a strange hidden-charm
  $\Xi_c\bar D^*$ molecular pentaquark?},'' {\em Phys. Rev. D}, vol.~103,
  no.~5, p.~054007, 2021.

\bibitem{Chen:2020uif}
H.-X. Chen, W.~Chen, X.~Liu, and X.-H. Liu, ``{Establishing the first
  hidden-charm pentaquark with strangeness},'' {\em Eur. Phys. J. C}, vol.~81,
  no.~5, p.~409, 2021.

\bibitem{Wu:2021caw}
Q.~Wu, D.-Y. Chen, and R.~Ji, ``{Production of $P_{cs}(4459)$ from $\Xi_b$
  Decay},'' {\em Chin. Phys. Lett.}, vol.~38, no.~7, p.~071301, 2021.

\bibitem{Lu:2021irg}
J.-X. Lu, M.-Z. Liu, R.-X. Shi, and L.-S. Geng, ``{Understanding Pcs(4459) as a
  hadronic molecule in the $\Xi_b^-\to J/\psi\Lambda K^-$ decay},'' {\em Phys.
  Rev. D}, vol.~104, no.~3, p.~034022, 2021.

\bibitem{Zhu:2021lhd}
J.-T. Zhu, L.-Q. Song, and J.~He, ``{$P_{cs}(4459)$ and other possible
  molecular states from $\Xi_{c}^{(*)}\bar{D}^{(*)}$ and $\Xi'_c\bar{D}^{(*)}$
  interactions},'' {\em Phys. Rev. D}, vol.~103, no.~7, p.~074007, 2021.

\bibitem{Malabarba:2021taj}
B.~B. Malabarba, K.~P. Khemchandani, and A.~M. Torres, ``{$N^*$ states with
  hidden charm and a three-body nature},'' {\em Eur. Phys. J. A}, vol.~58,
  no.~2, p.~33, 2022.

\bibitem{Yalikun:2021bfm}
N.~Yalikun, Y.-H. Lin, F.-K. Guo, Y.~Kamiya, and B.-S. Zou, ``{Coupled-channel
  effects of the $\Sigma_c ^*D^{*-}\Lambda_(2595)D^-$ system and molecular
  nature of the $P_c$ pentaquark states from one-boson exchange model},'' {\em
  Phys. Rev. D}, vol.~104, no.~9, p.~094039, 2021.

\bibitem{Zhu:2015bba}
R.~Zhu and C.-F. Qiao, ``{Pentaquark states in a diquark\textendash{}triquark
  model},'' {\em Phys. Lett. B}, vol.~756, pp.~259--264, 2016.

\bibitem{Ali:2019npk}
A.~Ali and A.~Y. Parkhomenko, ``{Interpretation of the narrow $J/\psi p$ Peaks
  in $\Lambda_b \to J/\psi p K^-$ decay in the compact diquark model},'' {\em
  Phys. Lett. B}, vol.~793, pp.~365--371, 2019.

\bibitem{Shi:2021wyt}
P.-P. Shi, F.~Huang, and W.-L. Wang, ``{Hidden charm pentaquark states in a
  diquark model},'' {\em Eur. Phys. J. A}, vol.~57, no.~7, p.~237, 2021.

\bibitem{Azizi:2021utt}
K.~Azizi, Y.~Sarac, and H.~Sundu, ``{Investigation of $P_{cs}(4459)^0$
  pentaquark via its strong decay to $\Lambda J/\Psi$},'' {\em Phys. Rev. D},
  vol.~103, no.~9, p.~094033, 2021.

\bibitem{Wang:2019got}
Z.-G. Wang, ``{Analysis of the $P_c(4312)$, $P_c(4440)$, $P_c(4457)$ and
  related hidden-charm pentaquark states with QCD sum rules},'' {\em Int. J.
  Mod. Phys. A}, vol.~35, no.~01, p.~2050003, 2020.

\bibitem{Wang:2020eep}
Z.-G. Wang, ``{Analysis of the $P_{cs}(4459)$ as the hidden-charm pentaquark
  state with QCD sum rules},'' {\em Int. J. Mod. Phys. A}, vol.~36, no.~10,
  p.~2150071, 2021.

\end{thebibliography}
\bibliographystyle{ieeetr}
\end{document}